\documentclass[aps,prl,reprint,floatfix]{revtex4-1}
\usepackage{amsmath}
\usepackage{graphicx}

\usepackage[scanall]{psfrag}

\newcommand{\abs}[1]{\left| #1 \right|}

\begin{document}

\title{Universality of striped morphologies}%

\author{E. Edlund}%
 \email{erik.edlund@chalmers.se} 
\author{M. \surname{Nilsson Jacobi}}%
 \email{mjacobi@chalmers.se} 
\affiliation{Complex Systems Group, Department of  Energy and Environment, Chalmers University of Technology, SE-41296 G\"oteborg, Sweden}
\date{\today}%

\begin{abstract}
We present a method for predicting the low-temperature behavior of spherical and Ising spin models with isotropic potentials. For the spherical model the characteristic length scales of the ground states are exactly determined but the morphology is shown to be degenerate with checkerboard patterns, stripes and more complex morphologies having identical energy. For the Ising models we show that the discretization breaks the degeneracy causing striped morphologies to be energetically favored and therefore they arise universally as ground states to potentials whose Hankel transforms have nontrivial minima.
\end{abstract}
\maketitle


The study of pattern formation in simple systems has received much attention the last  20 years. Not only are the pictures visually arresting, producing reviews with considerable artistic qualities~\cite{seul_domain_1995,bowman_natural_1998}, but spatially inhomogeneous phases present difficulties for standard theoretical methods, calling for new principles to describe the physics of systems exhibiting them~\cite{emery_stripe_1999}. Whatever weight can be assigned to the former as an explanation for the interest, the latter is justification enough, especially as the experimental evidence of stripes, spots and checkerboards in practically important materials abound. Examples include lipid monolayers~\cite{keller_stripe_1999}, adsorbates on metals~\cite{kern_long-range_1991}, and various magnetic fluids~\cite{rosensweig_labyrinthine_1983,seul_evolution_1992}. Striped  phases are also hypothesized to play a role in the high-temperature superconductivity of transition metal oxides~\cite{tranquada_evidence_1995,orenstein_advances_2000}.

Many experimental systems displaying heterogeneous patterns involve a competition between long- and short-range interactions~\cite{debell_dipolar_2000} and most theoretical work concentrate on specific examples of such interactions~\cite{garel_phase_1982,lw_study_1994,grousson_phase_2000,stoycheva_stripe_2000,giuliani_ising_2006}, e.g.\ spin models with Hamiltonians on the form
\begin{equation}
\label{hamiltonian1}
H=K \sum_j s_j^2 - L \sum_{\langle i,j \rangle} s_i s_j + \frac{Q}{2}\sum_{i\neq j} \frac{s_i s_j}{r_{ij}^\alpha},
\end{equation}
where the spins typically represent some coarse-grained feature of the system of interest, for example local charge density in a Mott insulator~\cite{emery_stripe_1999} or phases in a Langmuir film~\cite{weis_two-dimensional_1984}. 
However, the qualitative success of such models may have little to do with the underlying physics as noted by Zaanen in the context of Mott insulators~\cite{zaanen_current_1998}. Indeed, the same general behavior can be observed in models with for example only short-ranged, purely repulsive forces~\cite{malescio_stripe_2003}.
An explanation for the universality of striped morphologies must therefore be independent of specific details of the involved forces. The aim of this Letter is to present such a general treatment. As expected our method shows that stripes appear naturally for large classes of models, but the added generality also leads to new tools allowing us to design potentials with desired properties.

Here we study a generic Hamiltonian with isotropic pairwise interactions
\begin{equation}
\label{hamiltonian}
H=\sum_{ij}^N V_{ij} s_i s_j
\end{equation}
where  $V_{ij} = V ( \abs{i - j} )$ is a matrix representation of the potential that only depends on the distance between spins $i$ and $j$, here denoted $\abs{i - j}$, with \eqref{hamiltonian1} as a special case. Depending on considerations regarding experimental fit or theoretical ease, one may take the spins in \eqref{hamiltonian} to assume continuous values with the restriction $\sum_i s_i^2=N$, corresponding to a spherical model~\cite{baxter_exactly_1982}, or take values from some finite set, where $s\in\{\pm1\}$ and $s\in\{0,\pm1\} $ are the most common choices, equivalent to different Ising models~\cite{lw_study_1994, giuliani_ising_2006}.

Consider now the spherical model. Due to the quadratic form of the Hamiltonian~\eqref{hamiltonian} and the constraint, ${\bf s}^T V {\bf s}$ and ${\bf s}^T {\bf s} = N$ in matrix notation, the ground state  is given by an eigenvector of the interaction matrix $V$ corresponding to the lowest (energy) eigenvalue~\cite{lay_linear_2003}.
The central observation for our analysis is the existence of a common basis of eigenvectors for all radial potentials, namely the Fourier basis.
\begin{figure}[bp]
\includegraphics[width=0.28\textwidth]{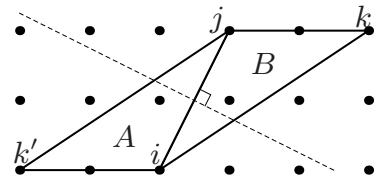}
\caption{ \label{triag} Reflection symmetry of triangles on a lattice causes the interaction matrices from any two radial potentials to commute.}
\end{figure}
To prove this we start by recalling the fact that if two matrices commute, then it is possible to find a set of eigenvectors that simultaneously diagonalize them~\cite{horn_matrix_1990}. Consider the commutator for two interaction matrices $V$ and $W$: 
 \begin{equation}
	 \sum _{k'} V_{ik'} W_{k'j} - \sum_{k} W_{i k} V_{k j} .
\label{commute}
\end{equation}
Each term, $V_{ik'} W _{k'j}$, in the first sum  can be represented by a triangle, A in Fig. \ref{triag}. Assuming that the lattice is periodic or infinite, there will for each such triangle exist a unique triangle B, constructed as a reflection of A in the line equidistant from point $i$ and $j$ (dashed in the figure), corresponding to the term $W_{ik} V _{k j}$ in the second sum. From the reflection symmetry and the pure radial dependence of the potential it follows that $V_{ik'} W _{k'j} - W_{ik} V _{k j} = 0$, which proves that $V$ and $W$ commute.

It now suffices to find a set of eigenvectors for a particular potential. Perhaps the simplest choice is a nearest neighbor interaction, $V_{ij}=1$ if $\abs{i - j}  =1$ and $V_{ij}=0$  if $\abs{i - j} >1$. If we in addition make an appropriate choice of self-interaction, which only shifts the eigenvalues and do not affect the eigenvectors, $V_{ii} = -2d$ where $d$ is the lattice dimension,  $V_{ij}$ becomes a discrete finite difference Laplacian on the lattice. It is well known that both the discrete and continuous Laplacian have harmonic eigenfunctions, e.g. $f_{\vec{k}}(\vec{x}) = C_{\vec{k}} \prod_i^d \cos{(2\pi k_i x_i/L + \phi _i )}$ which is an orthogonal eigenbasis in $d$ dimensions when $\vec{k}$ goes over all distances on the reciprocal lattice, $L$ is the linear size of the lattice, $\phi _i =   \pm \pi/4$ and $C_{\vec{k}}$ is an appropriate normalization constant. We have thus shown that all interaction matrices have a Fourier eigenbasis. An alternative, more direct but for our purposes less illustrative, argument for the common Fourier basis is to note that the structure of $V$ implies that it is a so called circulant matrix~\cite{davis_circulant_1994}, for which the result is known in the signal processing literature. That the Fourier base effectively diagonalize the Hamiltonian in the spherical model with translationally  invariant interactions has also been pointed out by Nussinov~\cite{nussinov_commensurate_2001}.

This result has two important consequences. First, it helps us to understand why systems with different interactions are expected to have similar ground states. Second, knowledge of the universal eigenbasis allows us to compute the energy spectrum for any particular system using a linear transform of the potential. From~\eqref{hamiltonian} it follows that the energy per spin for a harmonic eigenfunction with wave vector $\vec{k}$ are given by $ E(\vec{k})=\frac{1}{N}\sum_{\vec{x},\vec{y}}V(\abs{\vec{x}-\vec{y}}) f_{\vec{k}}(\vec{x}) f_{\vec{k}}(\vec{y})$. Using various trigonometric identities and the radial structure of $V$ this expression can be reduced to a radial Fourier transform
\begin{equation}
\label{discrete}
E(\vec{k}) = \sum_{\vec{r}} V(\abs{\vec{r}}) \prod_{i=1}^d \cos{(2\pi k_i r_i/L)}
\end{equation}
where the sum goes over all distances $\vec{r}$ on the lattice.  Note that the energy of a configuration can be computed through a fast Fourier transform over the lattice.

\begin{figure}[thb]
\includegraphics[width=0.34\textwidth]{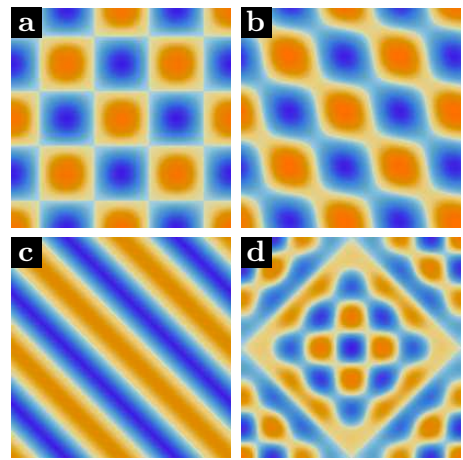}
\caption{
\label{sphericalGround} Examples of eigenmodes of the two dimensional spherical model. {\bf a}-{\bf c}, Any phase shift of a ground state is a new ground state, so checkerboards, stripes and everything between can be produced by the same model. Linear combinations of $f_{(2,2)}$ with different phase shifts are shown, all having the same energy. {\bf d}, Exchanging the elements of $\vec{k}$ gives a new ground state and linear combinations of them give rise to complex morphologies. Shown is $\frac{1}{2} f_{(3,4)} + \frac{1}{2} f_{(4,3)}$. 
}
\end{figure}
The ground state of the spherical model is the eigenvector $f_{\vec{k} }$ corresponding to the minimum of $E(\vec{k})$. The simplest ground state patterns in two dimensions are checkerboards and stripes with the corresponding wavelength, exemplified in Fig.~\ref{sphericalGround}a and~c. Further, the subspace of the eigenbasis corresponding to the minimum can contain two kinds of degeneracies. First, any change of the phases $\phi _i$ leaves the energy invariant. In two dimensions this means that anything between checkerboards and stripes can be produced, as illustrated in Fig.~\ref{sphericalGround}a-c. Second, the energy is similarly unaffected by arbitrary permutations of the elements of $\vec{k}$, reflecting that the energy only depends on the magnitude of the wave vector (seen most clearly in the continuous limit~\eqref{cont}). Linear combinations of vectors with different permutations give rise to complex morphologies, exemplified  in Fig.~\ref{sphericalGround}d.

\begin{figure*}[htb]
\includegraphics[width=0.92\textwidth]{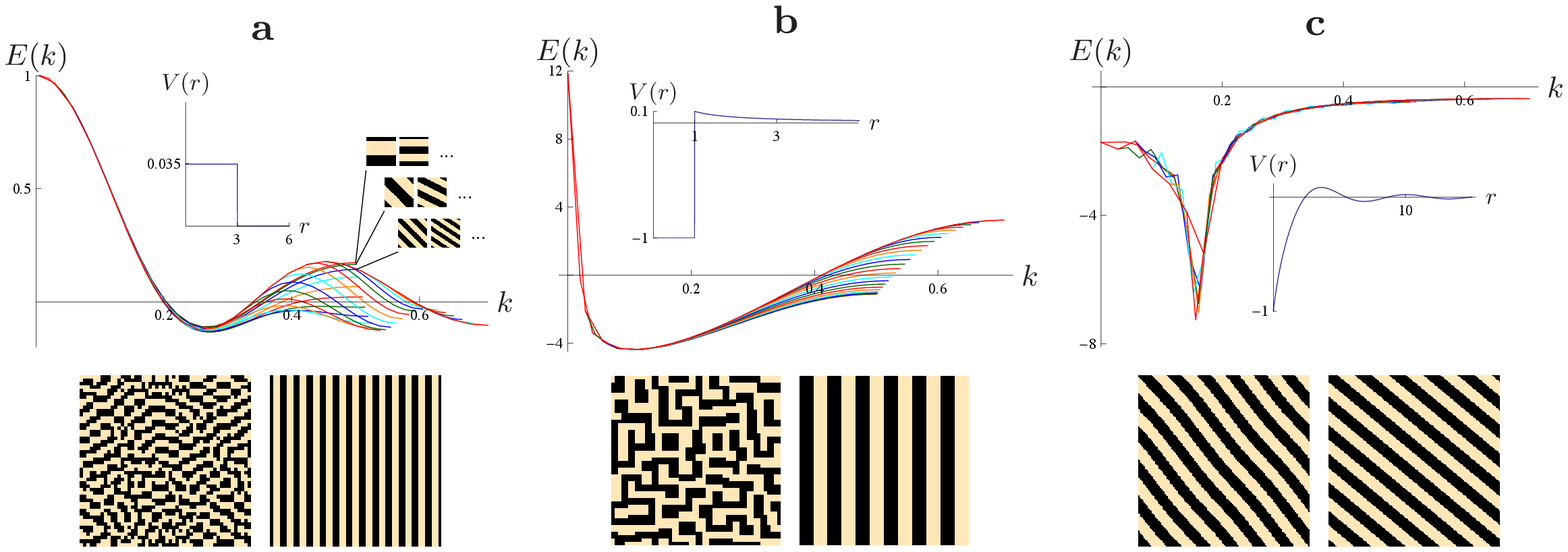}
\caption{ \label{potentials}
Predicting ground states in different two dimensional Ising models. 
(Top) Potentials (inset) and their energy spectra in $| \vec{k} |$-space from the transform \eqref{discrete}. {\bf a}, a purely repulsive potential~\cite{malescio_stripe_2003}, {\bf b}, two competing interactions~\cite{lw_study_1994} (see equation~\eqref{hamiltonian1}),  and {\bf c}, an RKKY-like interaction~\cite{fischer_spin_1993}. For small $k$ (long wavelengths) the spectra only depend on the magnitude of $\vec{k}$, but for large $k$ (short wavelengths) lattice effects breaks the independence on the direction of $\vec{k}$ and  $E(| \vec{k} |)$ becomes multivalued. To illustrate this, we link series with constant $k_y$ with lines and in {\bf a} examples of such states are shown.
(Below) Local minima as arrived by through Monte Carlo annealing as well as ground states of Ising spin-1/2 models with corresponding potentials.}
\end{figure*}

The eigenmode analysis is exact for the spherical model but also has implications for the discrete Ising models. It is not directly applicable as in general an eigenvector of the interaction matrix cannot be constructed in the restricted discrete space of the Ising spins. However, continuous eigenvectors are often used to approximate solutions to discrete optimization problems, for example graph coloring~\cite{aspvallF} and partitioning networks into modules with minimal intra-connectivity~\cite{fiedler73, newman06}. Here we use the same strategy to predict ground states for Ising spin-$1/2$ models with corresponding potentials by mapping the spins in the spherical model to $-1$ or $+1$ depending on their sign: $\hat{f} _{\vec{k}}(\vec{x})  = \mbox{sign} ( f_{\vec{k}}(\vec{x}) )$.%
%
~\footnote{This discretization of the Fourier modes to approximate the ground states of Ising models was also discussed in~\cite{nussinov_commensurate_2001} where it is argued that striped ground states should be energetically favored if the minimum in the energy spectrum is sharp. The argument we present does not have this assumption which is important as many commonly used potentials have broad energy minima and striped ground states, see e.g.\ Fig.~\ref{potentials} {\bf a} and {\bf b}.}
%
The discretization breaks the energy degeneracy and stripes become energetically favorable compared to checkerboards and more complex patterns. 
To see why we note that in each group of degenerate eigenmodes,  with wavelength $| \vec{k}|$, there exist linear combinations that produce stripes, for example $\cos (  \vec{k} \cdot \vec{x} )$. The error introduced by the discretization, $\|  \hat{f} _{\vec{k}}(\vec{x}) - f _{\vec{k}}(\vec{x}) \|_2$ with the standard $L^2$ norm,  always increases the energy in the discrete configurations when compared to the continuous ground state. Due to the $\pm$-symmetry of the harmonic functions, the difference between $\hat{f}_{\vec{k}}(\vec{x} )$ and $f_{\vec{k}}(\vec{x} )$ (with appropriate scaling) is largest in regions where the continuous function is close to $0$, i.e.
 at the interface between $+$ and $-$ regions. From this argument it follows that the error
 tends to be smallest for the striped eigenmode since the interface  is minimized (assuming that the width of the stripes is large compared to the lattice spacing).%
%
%

There are two exceptions when the ground state does not have stripes. Energy spectra with minimum at the boundary produce ground states that are either a uniform ferromagnet (the zero frequency mode) or a checkerboard pattern (the highest frequency mode allowed on the lattice) associated with an anti-ferromagnet. These two cases can be viewed as degenerate cases of stripes with infinite respective infinitesimal width.

\begin{figure*}[htb]
\includegraphics[width=0.91\textwidth]{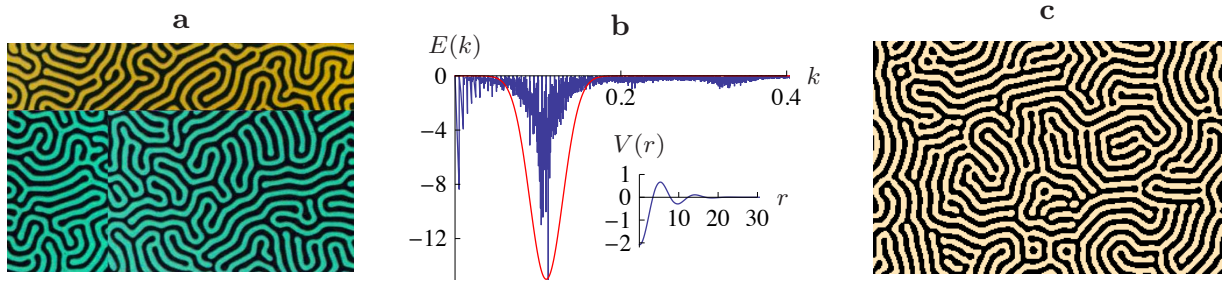}
\caption{ \label{reverseEng} Designing interactions to imitate observed striped patterns. 
{\bf a}, Stripes in a ferrofluid confined between two glass plates in a magnetic field [from~\cite{rosensweig_labyrinthine_1983,seul_domain_1995}, reprinted with permission from AAAS and Elsevier] 
{\bf b}, the (negative) radial power spectrum (blue) of previous picture together with a gaussian (red), (inset) a pair potential $V(r)$ constructed as the (inverse) transform~\eqref{discrete} of said gaussian and
{\bf c}, metastable state of an Ising spin-$1/2$ model with potential $V(r)$. Note that the chosen energy spectrum is not unique. Many potentials having a spectrum minimized at the same wavelength will show similar low-temperature behavior.
}
\end{figure*}

In Fig.~\ref{potentials} some examples of  Ising spin-$1/2$ models with different potentials are shown together with their energy spectra in $|\vec{k}|$-space, examples of local minima%
~\footnote{Local minima of the spin systems are found by a standard Metropolis algorithm on lattices with periodic boundaries. The results shown in Fig.~\ref{potentials} where generated by simulating 50$\times$50-lattices (in a and b) or a 150$\times$150-lattice (in c) with the potentials shown at $T=0.1$ for $2.5\times10^5$ trial flips and then with $T=0$ until convergence. In Fig.~\ref{reverseEng} a 300$\times$200-lattice with $6\times10^5$ trial flips at $T=0.2$ was used.}
and their ground states. First is a short-ranged, purely repulsive potential related to the model studied in~\cite{malescio_stripe_2003}. Second is a nearest neighbor ferromagnet with long-range repulsive Coulomb interaction on the form~\eqref{hamiltonian1} from~\cite{lw_study_1994}. Last is an attenuated Bessel function, $J_0(r)/(r+1)$, chosen for its similarity to the RKKY interaction in spin glasses~\cite{fischer_spin_1993}. We see that, while the potentials are qualitatively very different, the ground states are defined only by the minima in the energy spectrum, i.e. by a single length scale. Through rescaling, the potentials can be adjusted to have identical ground states. This illustrates how little observing striped behavior tells us about the interactions in a system. The local minima do however show a qualitative difference between the potentials in {\bf a} and {\bf b} and the RKKY-like potential in {\bf c}, probably related to the difference in localization in energy space. 

 Equation \eqref{discrete}  also has implications for molecular self-assembly. The Fourier basis in the transformation is orthogonal and can be inverted to find the potential corresponding to a given energy spectrum. This allows us to design, from an observed striped state, families of potentials that generate similar patterns at low-temperature by identifying the dominant wavelength and invert an energy spectrum with a minimum at this wavelength.
 A demonstration of the procedure is shown in Fig.~\ref{reverseEng}: the Fourier power spectrum of a pixelised image of a metastable state in an experimental system~\cite{rosensweig_labyrinthine_1983} was calculated;  an energy spectrum was constructed with gaussian minimum at the same wavelength as the experimental system; and finally the corresponding potential was found using the inverse transform of~\eqref{discrete}. The constructed system has striped metastable states similar to those found in the experimental system. We conclude that it is relatively easy to construct families of potentials with desired metastable striped morphologies.

In the continuous limit the transformation \eqref{discrete} becomes a Hankel transform, in two dimensions defined as
\begin{equation}
\label{cont}
E ( \vec{k} ) = 2\pi \int _{0} ^{\infty} r dr V(r) J_0 ( 2\pi | \vec{k} | r ) 
\end{equation}
where $J_0$ is a Bessel function of the first kind. For the general expression in higher dimensions, see~\cite{folland_fourier_2009}. Note that in the continuum limit the energy only depends on the magnitude of the wavevector since the effects of the principal lattice directions disappear. As noted in Fig.~\ref{potentials}, this independence holds true on the lattice as well for small wavevectors and sufficiently long-range interactions. Equation~\eqref{cont} allows us to use the analytical properties of the Hankel transform to qualitatively understand for example why the Bessel function of Fig.~\ref{potentials}c has such a sharp spectrum: the Hankel transform of a Bessel function is a Dirac delta function.


In summary we have shown that the energy spectrum of spherical spin systems with isotropic interactions can be derived directly from the Fourier transform of the potential. Due to a degeneracy in the energy eigenstates the spherical model has ground states with various patterns such as stripes, checkerboards, and more complicated morphologies. In discrete spin models the degeneracy is broken leading to striped ground states being energetically favored. We suggest that this can offer a generic explanation to why striped patterns are so frequently observed in various experimental and natural systems. 

\begin{acknowledgments}
The authors would like to thank Olle H\"aggstr\"om for pointing out how purely repulsive potentials can give rise to striped ground states.
\end{acknowledgments}

\bibliographystyle{apsrev4-1}
\bibliography{StripedMorphologies}

\end{document}